\RequirePackage{fix-cm}
\documentclass[twocolumn,natbib]{svjour3}          % twocolumn
\smartqed  % flush right qed marks, e.g. at end of proof
\usepackage{graphicx}
\usepackage{epstopdf}
\usepackage{float}
\usepackage[caption=false]{subfig}
\usepackage{multirow}

\journalname{}
\begin{document}\sloppy

\title{Numerical analysis of quasi-static fracture in functionally graded materials}

\titlerunning{Numerical analysis of quasi-static fracture in FGMs}        % if too long for running head

\author{E. Mart\'{\i}nez-Pa\~neda         \and
        R. Gallego %etc.
}

%\authorrunning{Short form of author list} % if too long for running head

\institute{E. Mart\'{\i}nez-Pa\~neda \at
              Department of Construction and Manufacturing Engineering\\
              University of Oviedo, Gij\' on 33203, Spain\\
              \email{martinezemilio@uniovi.es}           %  \\
%             \emph{Present address:} of F. Author  %  if needed
           \and
           R. Gallego \at
             Department of Structural Mechanics and Hydraulic \\
             Engineering, Advanced School of Civil Engineering\\ 
             University of Granada, Granada 18071, Spain
}

\date{Received: date / Accepted: date}
% The correct dates will be entered by the editor

\maketitle

\begin{abstract}
This work investigates the existing capabilities and limitations in numerical modeling of fracture problems in functionally graded materials (FGMs) by means of the well-known finite element code ABAQUS. Quasi-static crack initiation and growth in planar FGMs is evaluated. Computational results of fracture parameters are compared to experimental results and good agreement is obtained. The importance of the numerical fit of the elastic properties in the FE model is analyzed in depth by means of a sensitivity study and a novel method is presented. Several key computational issues derived from the continuous change of the material properties are also addressed and the source code of a user subroutine USDFLD is provided in the Appendix for an effective implementation of the property variation. The crack propagation path is calculated through the extended finite element method (X-FEM) and subsequently compared to available experimental data. Suitability of local fracture criteria to simulate crack trajectories in FGMs is discussed and a new crack propagation criterion is suggested.
\end{abstract}

\section{Introduction}
\label{intro}
Functionally Graded Materials (FGMs) are those whose composition and hence their properties vary gradually as a function of the position. Since their introduction by \cite{Kawasaki1987} in high temperature metal/ceramic aerospace components, FGMs have found a wide range of commercial applications including cutting tools, biomedical devices, optical fibers and wear resistant coatings \citep{Uemura2003}. In many of these applications, FGMs provide an attractive way for the designer to tailor the microstructure to specific operating conditions, while minimizing the difficulties associated with discrete material interfaces. Very often, however, fracture resistance constitutes the primary design criterion, and this fact has led to the development of a special branch of fracture mechanics devoted to the failure of this class of materials.

Until now, the fracture of an FGM under quasi-static loading, which is one of the predominant modes of material failure, has been investigated extensively \citep{Eischen1987,Jin1994,Erdogan1995}. The primary conclusion of these investigations is that the classical inverse square root singular nature of the stress field is preserved in FGMs, but the stress intensity factor (SIF) is influenced by the non-homogeneity of the material. Therefore, in a linear-elastic cracked FGM, SIFs play a significant role since they characterize the crack-tip stress and strain fields. The non-singular $T$-stress, which represents the stress parallel to crack faces, is another factor affecting the crack growth behavior \citep{Becker2001}.

The finite element method (FEM) has been widely used for fracture analyses of FGMs. \cite{Eischen1987} has evaluated mixed-mode SIFs by means of the path-independent $J_{k}^{*}$ integral. \cite{Bao1995} have investigated periodic cracking in graded ceramic/metal coatings. \cite{Gu1997} have evaluated SIFs using the standard $J$-integral. \cite{Bao1997} have studied delamination cracking in graded ceramic/metal substrate under mechanical and thermal loads. \cite{Anlas2000} have calculated SIFs by means of the path-independent $J_{1}^{*}$ integral. \cite{Marur2000} have investigated a crack normal to the material gradient by means of both the FEM and experiments. \cite{Dolbow2002} have calculated the SIFs through the extended finite element method (X-FEM). \cite{Kim2002} have evaluated mixed-mode SIFs by means of the path-independent $J_{k}^{*}$ integral, the modified crack closure (MCC) and the displacement correlation technique (DCT). The $T$-stress has also been computed by means of the FEM. \cite{Becker2001} studied $T$-stress and finite crack kinking in FGMs. \cite{Kim2003} used a unified approach of the interaction integral method for evaluating SIFs and $T$-stress in FGMs. \cite{Chapa2002,Chapa2002b} have also used the FEM to investigate crack kinking in graded composites. 

On the experimental side, the difficulty and cost of manufacturing large-size fracture specimens amenable to testing has led most investigators to develop model FGMs. Of particular interest in this research is a model FGM based on polyethylene 1\% carbon monoxide co-polymer (ECO), manufactured by selective exposure to ultraviolet (UV) irradiation \citep{Lambros1999}. These specimens have material characteristics mimicking ceramic-metal FGMs, i.e., stiffer and more brittle at one end, becoming gradually less stiff and more ductile at the other. Cracks in these FGM ECO specimens have been analyzed by \cite{Abanto-Bueno2006} in the experimental work that has been chosen to validate the present numerical analysis. The difficulty in performing this type of experiments has led many analysts to adopt numerical schemes and solve FGM-related fracture problems. Although boundary integral formulations have been used in some cases \citep{Zhang2003,Riveiro2013}, the FEM is by far the approach most commonly adopted.

This work evaluates the performance of numerical tools in the computational assessment of cracks in FGMs by means of the well-known ABAQUS finite element (FE) code. Computational results of fracture parameters (SIFs and $T$-stress) are compared with available experimental results and good agreement is obtained. The importance of the numerical fit of the elastic properties in the finite element model is analyzed by means of a sensitivity study and a new method is presented and evaluated. FEM capabilities in various key issues from the numerical point of view, such as the implementation of the property variation at the element level or the effect of the material gradation in the computation of fracture parameters, are examined in depth and, in order to overcome the existing limitations in commercial FE packages, the source code of a user subroutine USDFLD is provided and several improvements are suggested.
 
The crack propagation path is simulated through the X-FEM and a good agreement with the experimental results of \cite{Abanto-Bueno2006} is obtained. This is of particular interest since work previously reported in the literature on this subject is limited. Suitability of local crack propagation criteria to simulate crack trajectory in FGMs is discussed and a novel crack propagation criterion is proposed.

\section{Model formulation}
\label{sec:model}

\subsection{Specimen geometry and parameters}
\label{sec:model1}

The experimental results reported in this study are taken from those obtained by \cite{Abanto-Bueno2006}. Details of the experimental procedure are described in that work and in others related \citep{Lambros1999,Li2000}, so only the ones relevant to the current research will be described here. 

\cite{Abanto-Bueno2006} manufactured polymeric model FGMs based on selective ultraviolet (UV) irradiation of polyethylene cocarbon monoxide (ECO). ECO is a very ductile semicrystalline copolymer that undergoes accelerated mechanical degradation when exposed to UV light, so that by gradually irradiating a sheet of the material from one end to the other, a sample with continuous in-plane property gradation from stiff and brittle to more compliant and more ductile can be obtained. A very thin sheet of in-plane dimensions 300$\times$150 mm$^2$ was irradiated for times varying from 5 h to 300 h. Once irradiated, the sheet was divided parallel to the irradiation direction, and two samples of 150$\times$150 mm$^2$ were obtained. One of these was then cut perpendicularly to the irradiation direction into 15 strips of 10 mm width, which were used in uniaxial tension tests to measure the Young's modulus $E$, failure stress $\sigma_f$, and failure strain $\varepsilon_f$ as a function of distance along the ECO sheet. The remaining 150$\times$150 mm$^2$ sample from the original sheet was used to generate SENT fracture specimens. Therefore, although the material property variation was measured independently of the fracture experiments, both originate from the same manufacturing process.
 
\cite{Abanto-Bueno2006} monitored the near-tip field using the optical technique of digital image correlation (DIC). The DIC measured displacement field was then used to extract the fracture parameters by performing a least square minimization of the asymptotic expression of the displacement field in the vicinity of the crack tip; as in the case of an homogeneous material, but with material properties evaluated at the crack tip position \citep{Eischen1987}.

The testing protocol of \cite{Abanto-Bueno2006} included mixed mode fracture experiments on the base homogeneous material and various graded FGM samples. Mixed-mode fracture is inherent to FGMs since for a crack inclined to the property gradation direction, the stress state near the crack tip is mixed-mode irrespective of the far field loading. In order to validate and develop a complete numerical investigation of the fracture process of FGMs, Abanto-Bueno and Lambros experimental work is especially interesting because evaluates the three characteristic geometries of mixed-mode fracture in FGMs. Thereby, near-tip mixity can be attained either by asymmetric external loading, as in the homogeneous case, or by placing the notch at an angle to the direction of mechanical property variation, or by a combination of both. The effect of each of these cases was investigated using three specimens labeled here FGM I, II, and III. The geometry, dimensions and measured variation of local material properties of the three specimens are shown in Table 1 and Fig. 1.

\begin{table}[h]
\caption{Dimensions of the FGM specimens}
\centering
\begin{tabular}{c c c c c c} 
\hline
& H & W & h & a & $\varphi$\\
& (mm) & (mm) & (mm) & (mm) & (rad)\\
 \hline
 FGMI & 75 & 70 & 37.5 & 30 & $\pi /2$\\
 FGMII & 90 & 70 & 32 & 26 & $\pi /3$\\ 
 FGMIII & 90 & 70 & 32 & 25 & $\pi /3$\\
 \hline
\end{tabular}
\label{tab:Table1}
\end{table}

\begin{figure*}[!ht]
    \subfloat[\label{subfig-1:Fig1a}]{%
      \includegraphics[width=0.48\textwidth]{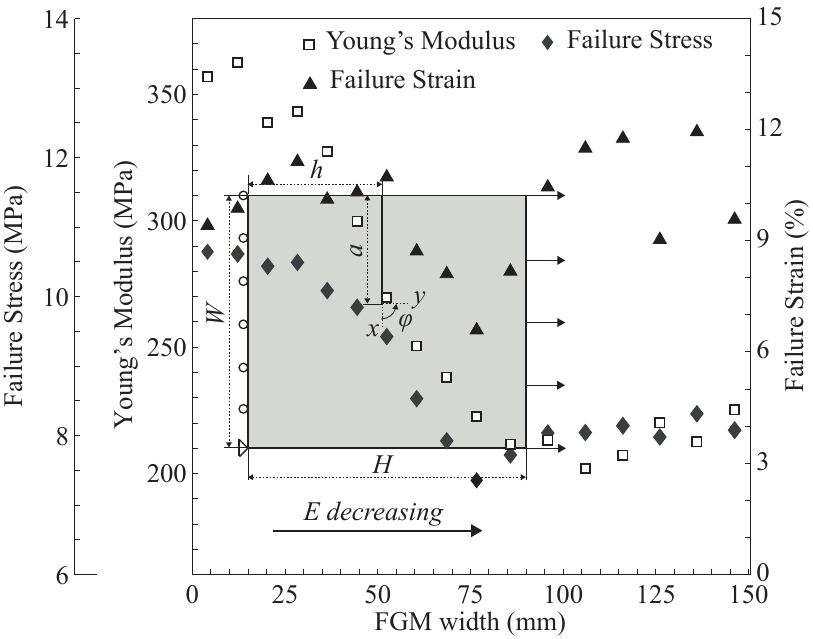}
    }
    \hfill
    \subfloat[\label{subfig-2:Fig1b}]{%
      \includegraphics[width=0.48\textwidth]{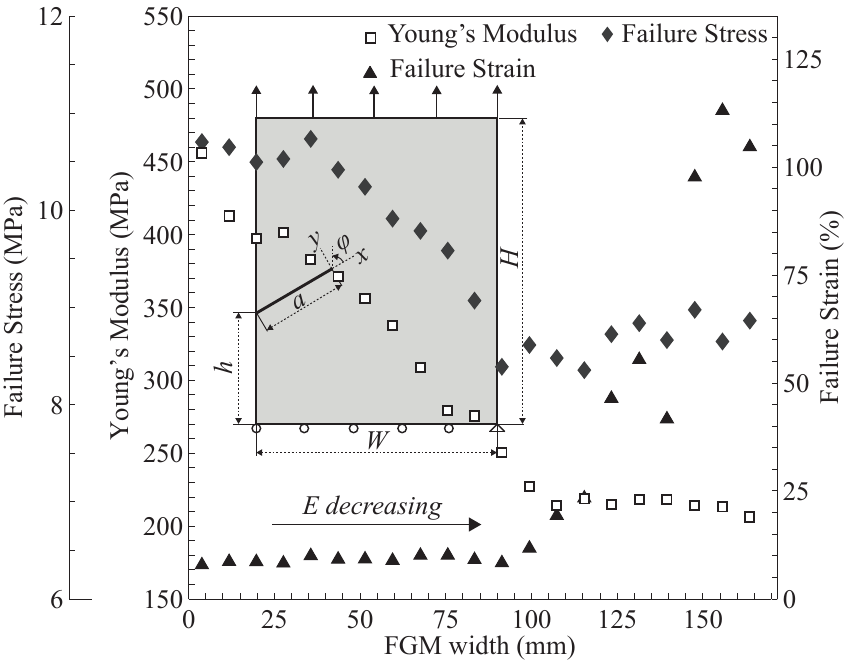}
    }
    
    \centering
    \subfloat[\label{subfig-2:Fig1c}]{%
      \includegraphics[width=0.48\textwidth]{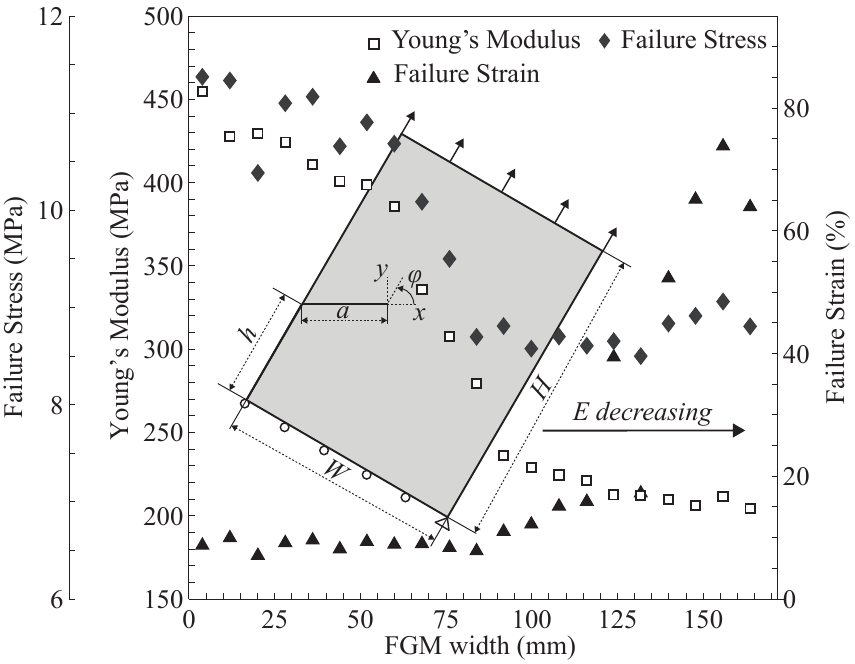}
    }
    \caption{Geometry, dimensions
and measured variation of local material properties as a function of the width of \textbf{(a)} FGMI, \textbf{(b)} FGMII and \textbf{(c)} FGMIII.}
    \label{fig:Fig1}
\end{figure*}

\subsection{Finite element model}
\label{sec:model2}

The simulations were developed in the latest version of \cite{ABAQUS}. The specimens dimensions correspond to those reported by \cite{Abanto-Bueno2006}. The sample thickness, common to all homogeneous and FGM cases, was 0.406 mm and thereby plane stress conditions are assumed. Mimicking the experimental procedure, loading is applied as a fixed vertical displacement along the upper edge of the specimen, the vertical displacement is constrained in the lower edge and, in order to remove rigid body motion, the horizontal displacement is also set to zero at the lower right hand corner. Given the fact that in both the homogeneous and graded cases the material used fails by crazing while showing very little shear yielding, linear-elastic behavior is assumed in this work as it was also assumed in the data analysis of \cite{Abanto-Bueno2006}.

\subsubsection{Application of material gradient}
\label{sec:model21}

The assignment of material properties must reflect the property distribution in the FGM specimen being simulated. However, almost all of the FE approaches mainly concentrate on homogeneous materials or piecewise homogeneous materials; specific FE formulations relating to nonhomogeneous materials with continuously varying properties are scarce. Nevertheless the inclusion of continual spatial variation of properties in the FE formulation does not entail a computational problem, as the stiffness matrix may be determined by averaging across each element. Material properties can vary between elements or between integration points. The former leads to a discontinuous step-type variation in properties. Assigning element properties individually, or dividing a structure into numerous areas and then assigning properties to areas \citep{Bao1995} may be inappropriate in failure analysis or crack path predictions, where local stress values may be of critical importance.

\cite{Santare2000} developed a formulation for graded elements, which automatically interpolate material properties within the element. These can substantially improve the solution quality based on the same mesh density, especially for higher-order graded elements. \cite{Kimb2002} have also investigated elements with an internal property gradient and reached similar conclusions. Their work differs in that the former samples the material properties directly at the Gauss integration points of the element, while the latter adopts a generalized isoparametric formulation.
 
\cite{Rousseau2000} developed a technique to assign spatially varying properties at integration points by defining properties as a function of temperature and providing the model with an initial temperature distribution that matches the elastic modulus variation desired. The assignment of a zero thermal expansion coefficient then eliminates unwanted thermal strains. This technique enjoys great popularity since it can be used in most of the commercial FE packages. However, it is not suitable for thermomechanical analyses and does not allow for differences between the gradient profiles of the Young's modulus and the Poisson's ratio. Furthermore, contrary to what is often assumed, Rousseau and Tippur's technique is not able to define a non-linear continuous variation of the elastic properties in most of the FE codes since, in order to obtain a consistent variation between mechanical and thermal strains, nodal temperature values are interpolated within the element through shape functions one order lower than those used for the displacements. In the case of ABAQUS, an average value of the temperature in the nodes is passed to the integration points when using linear elements and an approximate linear variation is assumed in quadratic elements. The former produces a step-type variation in the elastic properties (i.e., homogeneous elements) and the latter translates into a piecewise linear variation, regardless of the order of the function describing the elastic gradient. Therefore, one should be careful when assigning non-linear material property variations by means of Rousseau and Tippur's technique since, for coarse meshes and some gradient profiles, it can bring inaccuracies in the results. In ABAQUS this can be overcome by defining the gradation of properties through a user subroutine UMAT or USDFLD, since both are called at integration points. However, if a UMAT subroutine is used, the mechanical constitutive behavior of the material must also be programmed and hence, it is not possible to use the material models already implemented in ABAQUS. Consequently, the material gradient is implemented in this work through a USDFLD user subroutine. Material elastic properties are defined as a function of a field variable and its variation throughout the specimen is programmed in the subroutine. In addition, when computing the SIFs, the elastic properties in the crack tip must be defined and therefore a UFIELD subroutine is also embedded in the FORTRAN code in order to take into consideration as well the elastic properties variation at the nodes.

The source code of the subroutine is provided in the Appendix in order to allow other engineers to implement an \textit{effective} continuous variation of the material elastic properties without requiring programming efforts. Another option could be to use the research codes FGM-FRANC2D \citep{Kim2003b} or WARP3D \citep{WARP3D} since both include the gradation effect at the element level, based on the nodal-values approach \citep{Kim2002}. Both are freely distributed, open-source finite element codes with extended capabilities for fracture in FGMs, though the former is not yet available to the public and the latter does not have the capability to model plane stress conditions \citep{Walters2006}.

\subsubsection{Numerical fit of the material elastic properties}

The variation of composition in FGMs depends on the production technique \citep{Lambros1999,Butcher1999,Parameswaran2000} and generally, the property variation tends not to mirror that of composition. If the spatial composition profile is known, property variation may be predicted by means of theoretical mixing laws. Their use is frequent in composites \citep{Hashin1983} and has also been extended to FGMs \citep{Reiter1997,Gasik1998}. In these cases the assignment of the material property variation in the model is done straightforwardly, fitting the variation of the elastic properties through a function with the shape of the theoretical prediction, following the procedure mentioned in the previous paragraph. However, in such models predicted property variation is largely based in the assumed composite structure and, therefore, is usually limited in applicability and accuracy due to the geometric and micromechanical assumptions upon which the theoretical mixing laws are based.

Hence, material property variation is usually determined directly from experiment, being characterized by a sequence of experimental data as a function of the position, regardless of the form in which these data were obtained. Either by producing and testing individual homogeneous specimens with a range of compositions \citep{Carrillo-Heian2001,Jedamzik2000}, or by testing the graded material directly by means of indentation or ultrasonic techniques \citep{Krumova2001} or by cutting and testing small, effectively homogeneous, specimens from a larger graded sample \citep{Lambros1999,Rousseau2000,Butcher1999}, as in the case of the experimental work \citep{Abanto-Bueno2006} that serves as basis for the validation of the numerical model presented in this paper. To the authors' knowledge, the numerical fit of this experimental data, its implementation in the numerical model and its effect on the computational calculations have not received the attention of the scientific community. Mostly, the numerical fit is based on a general approximation of all the experimental data, either by assuming a linear change in the material properties \citep{Rousseau2000} or by means of a polynomial function through a least squares fit \citep{Oral2008b,Oral2008}. Following the criterion of the authors of the experimental study \citep{Oral2008b}, a fourth-order polynomial function was chosen to approximate the data (Fig. 2). But, as it can be seen in Fig. 2, it is impossible to completely remove the differences between the measured elastic properties and the polynomial curve fitting. And, even though no systematic study of the problem has been published yet, it is reasonable to expect that, in fracture analyses of FGMs, an accurate fit of the elastic properties near the crack tip could be overriding, due to the dependence of the fracture parameters' magnitude on the crack direction, property profile, crack-tip position and specimen geometry.

\begin{figure*}[!ht]
    \subfloat[\label{subfig-1:Fig2a}]{%
      \includegraphics[width=0.45\textwidth]{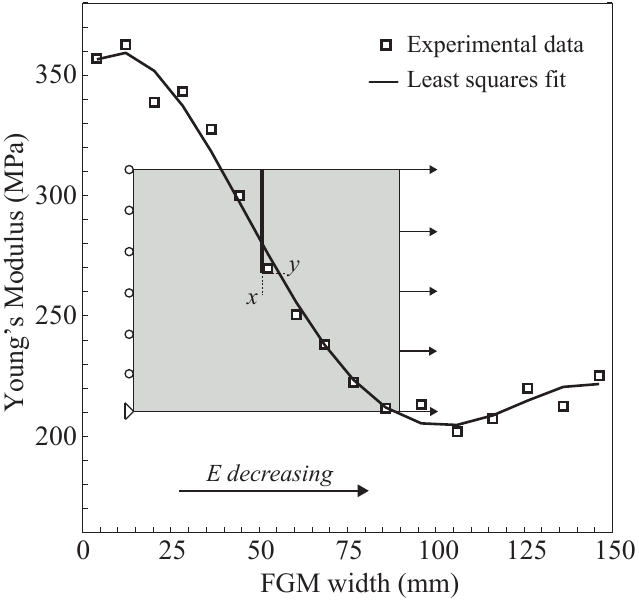}
    }
    \hfill
    \subfloat[\label{subfig-2:Fig2b}]{%
      \includegraphics[width=0.45\textwidth]{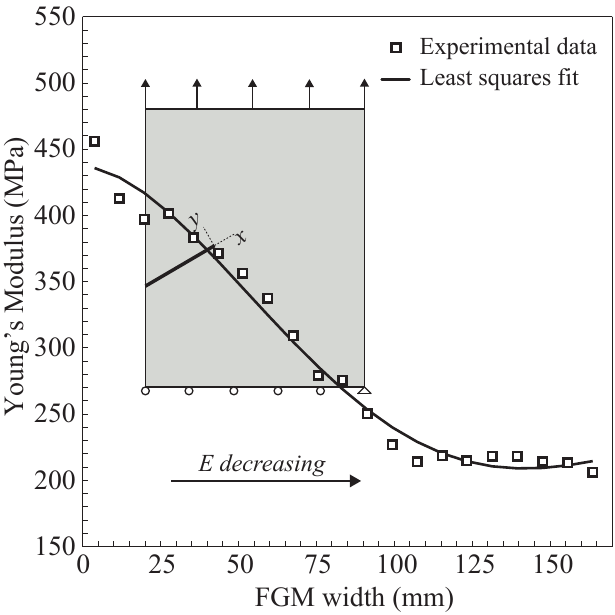}
    }
    
    \centering
    \subfloat[\label{subfig-2:Fig2c}]{%
      \includegraphics[width=0.45\textwidth]{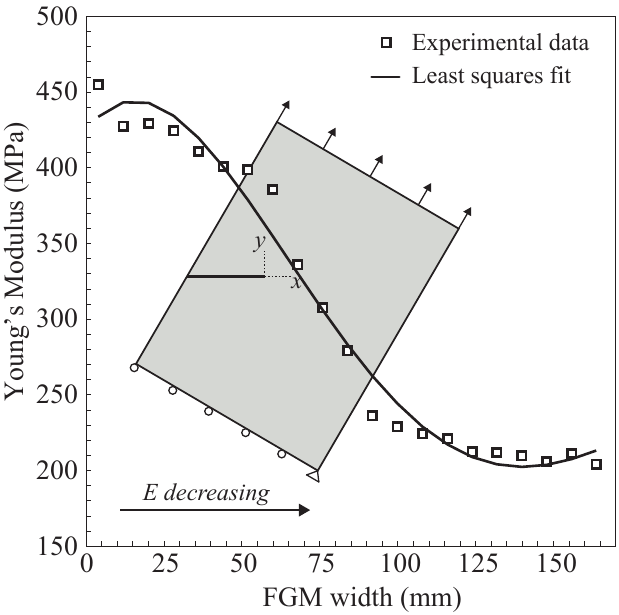}
    }
    \caption{Numerical fit of the material property variation of \textbf{(a)} FGMI, \textbf{(b)} FGMII and \textbf{(c)} FGMIII.}
    \label{fig:Fig2}
\end{figure*}

In order to rate the effect of the numerical fit of the experimental data in the calculation of the fracture parameters a complete sensitivity study is developed. First, the effect of an accurate fit of the local values of the elastic properties at the crack tip is analyzed. In order to do that, the value of the experimental point that characterizes the local elastic properties in the crack tip is modified a small amount (5\% - 10\%). For each of the FGM specimens evaluated, the new data points derived from these changes are shown in Table 2 and the corresponding polynomial curve fits can be seen in Fig. 3. SIFs and $T$-stress for each curve fit are calculated and compared.\\

\begin{figure*}[!ht]
    \subfloat[\label{subfig-1:Fig3a}]{%
      \includegraphics[width=0.48\textwidth]{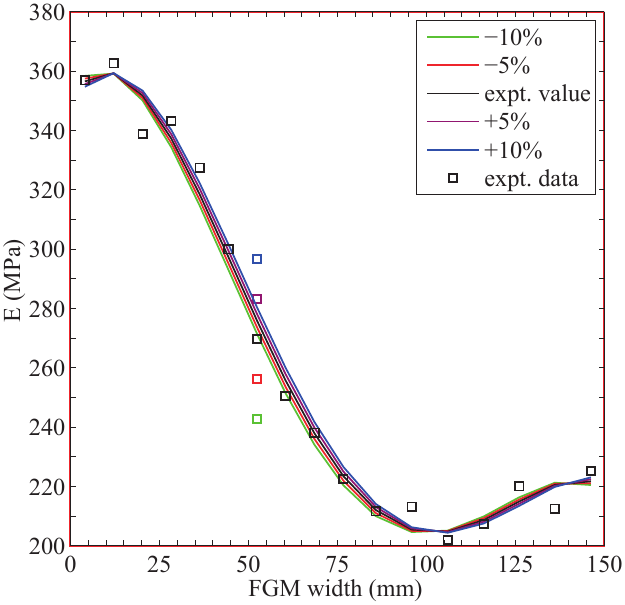}
    }
    \hfill
    \subfloat[\label{subfig-2:Fig3b}]{%
      \includegraphics[width=0.48\textwidth]{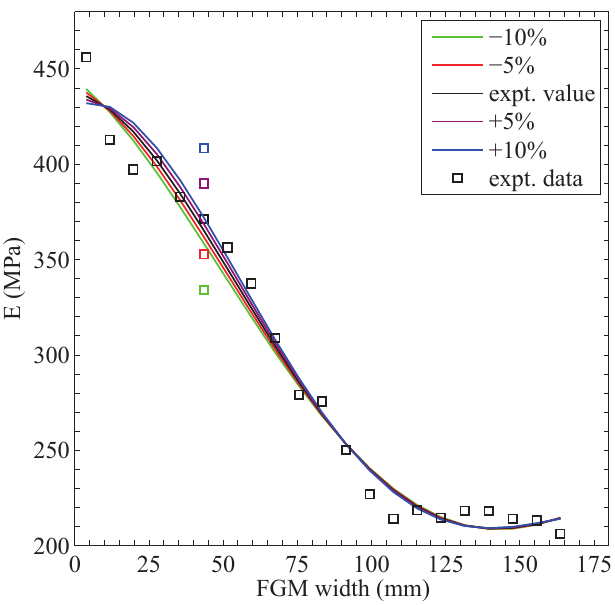}
    }
    
    \centering
    \subfloat[\label{subfig-2:Fig3c}]{%
      \includegraphics[width=0.48\textwidth]{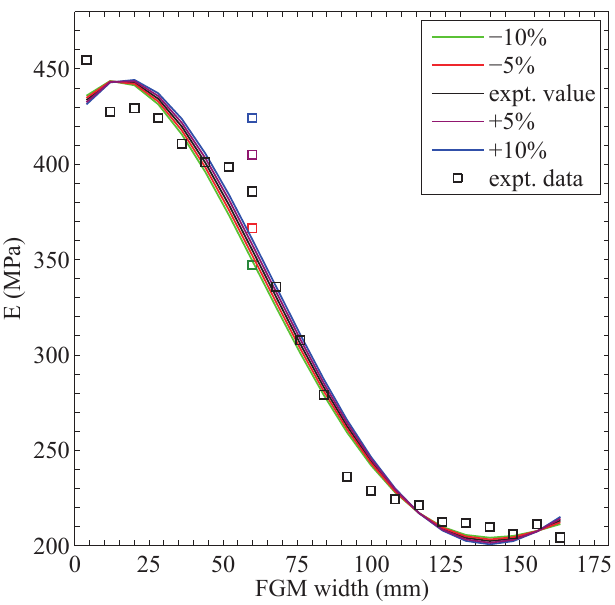}
    }
    \caption{Numerical fit of the material property variation taking into account the proposed amendments of \textbf{(a)} FGMI, \textbf{(b)} FGMII and \textbf{(c)} FGMIII.}
    \label{fig:Fig3}
\end{figure*}

\begin{table}[h]
\caption{New data points according to the proposed amendments for each specimen evaluated}
\centering
\begin{tabular}{c c c c} 
\hline \hline
& FGM I & FGM II & FGM III\\
 \hline
 -10\% & 242.72856 & 334.18539 & 347.14386\\
 -5\% & 256.21348 & 352.751245 & 366.42963\\ 
 Expt. data & 269.6984 & 371.3171 & 385.7154\\
 +5\% & 283.18332 & 389.882955 & 405.00117\\
 +10\% & 296.66824 & 408.4481 & 424.28694\\
 \hline \hline 
\end{tabular}
\label{tab:Table2}
\end{table}

Next, with the aim of evaluating the influence of an accurate fit of the material gradient profile in the computation of the SIFs and the $T$-stress, several polynomial curve fits of different orders are considered, containing all of them the experimental data characterizing the local property values in the crack tip, as shown in Fig. 4. Fracture parameters for each curve fit are calculated and compared. Based on the conclusions of the sensitivity analysis a new method is developed and evaluated.

\begin{figure*}[htbp]
    \subfloat[\label{subfig-1:Fig4a}]{%
      \includegraphics[width=0.42\textwidth]{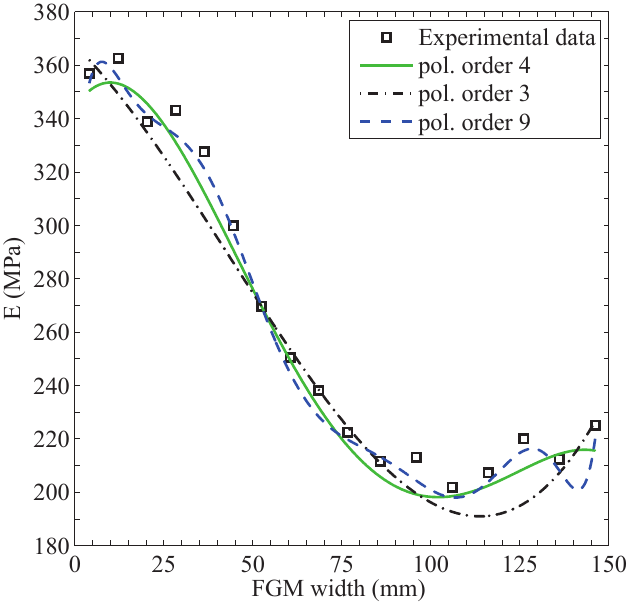}
    }
    \hfill
    \subfloat[\label{subfig-2:Fig4b}]{%
      \includegraphics[width=0.42\textwidth]{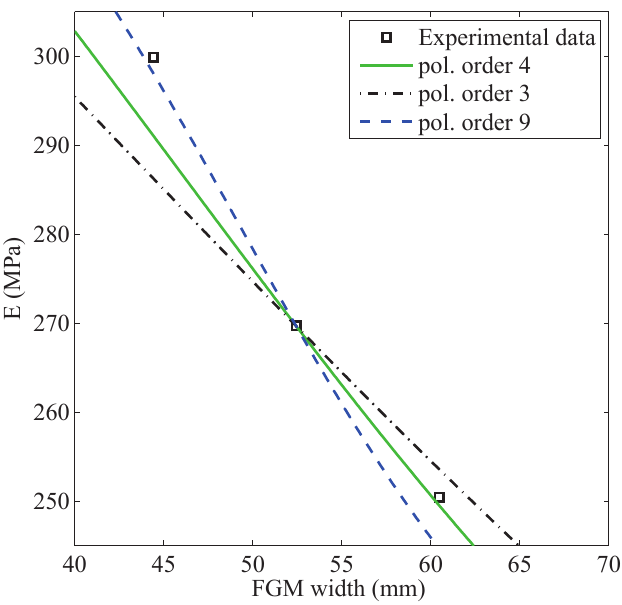}
    }
    
    \subfloat[\label{subfig-2:Fig4c}]{%
      \includegraphics[width=0.42\textwidth]{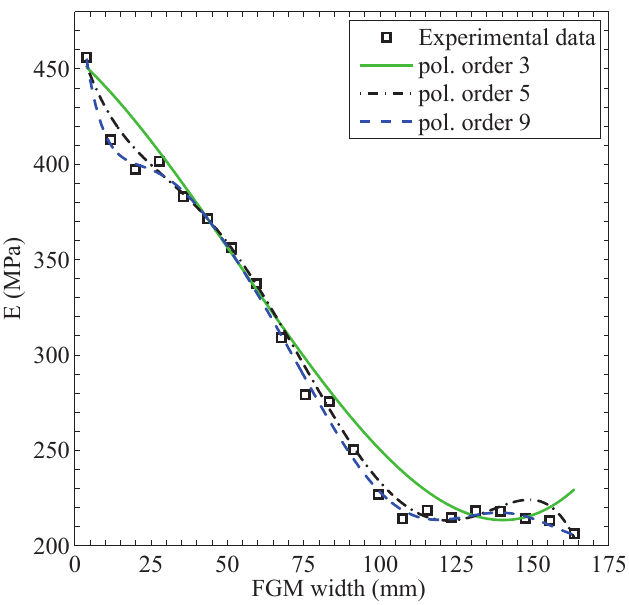}
    }
    \hfill
    \subfloat[\label{subfig-2:Fig4d}]{%
      \includegraphics[width=0.42\textwidth]{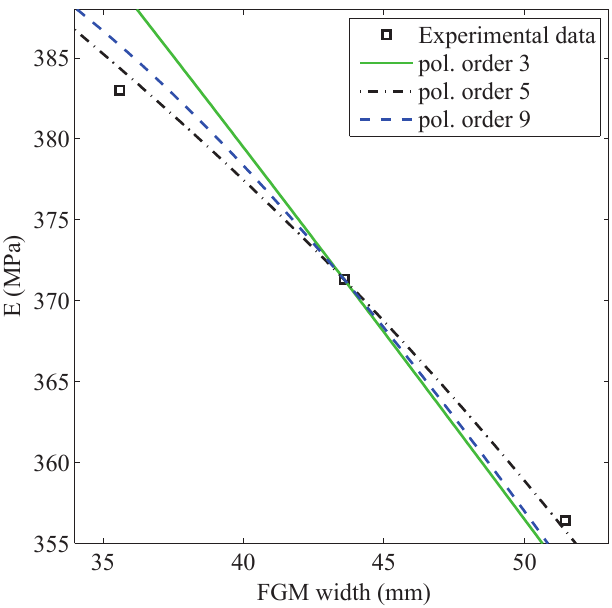}
    }
    
    \subfloat[\label{subfig-1:Fig4e}]{%
      \includegraphics[width=0.42\textwidth]{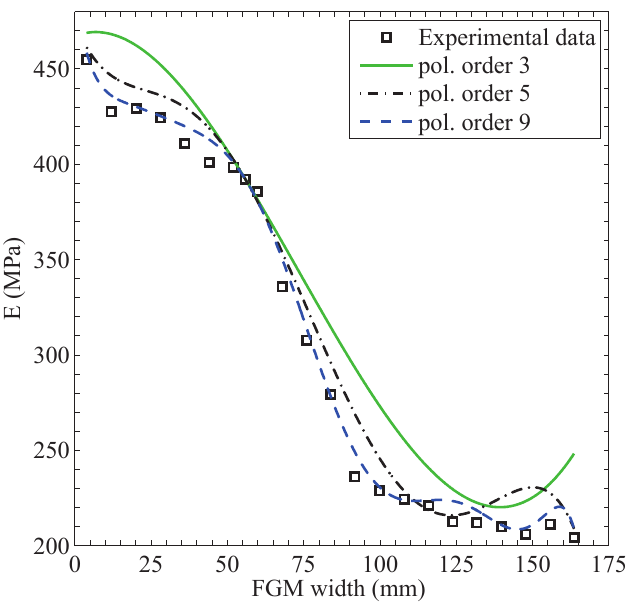}
    }
    \hfill
    \subfloat[\label{subfig-2:Fig4f}]{%
      \includegraphics[width=0.42\textwidth]{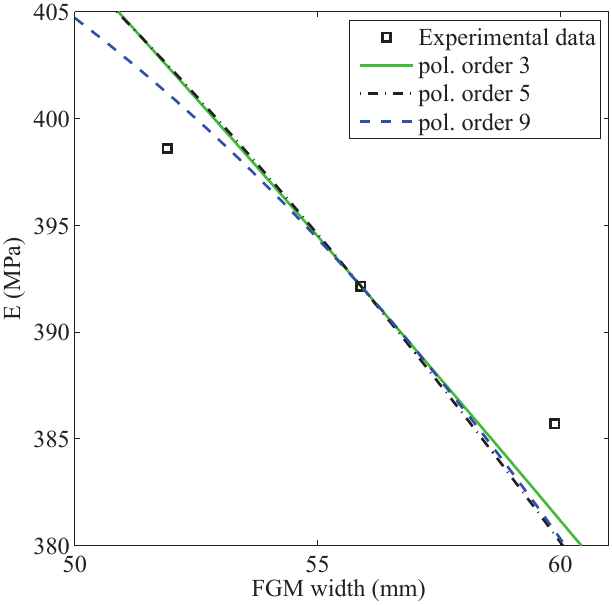}
    }
    \caption{Numerical fit of the experimental data by polynomial functions of different order of \textbf{(a)} FGMI, \textbf{(c)} FGMII and \textbf{(e)} FGMIII. Detail of the vicinity of the crack of \textbf{(b)} FGMI, \textbf{(d)} FGMII and \textbf{(f)} FGMIII.}
    \label{fig:Fig4}
\end{figure*}

\subsubsection{Calculation of fracture parameters}

By far the most common concern pertaining to linear elastic fracture mechanics analysis is the accurate prediction of SIFs in arbitrarily shaped cracked bodies. There are usually several ways to calculate fracture parameters once the stress and displacement fields have been obtained. In the displacement-based stress intensity factor computation techniques, the SIFs are obtained by extrapolating from the displacement ahead of the crack tip in the asymptotic expression. These methods have the advantage that almost no additional calculation is necessary, but they require a high degree of mesh refinement and often suffer from instability as the crack tip is approached \citep{Anderson2005}. Also, for the FGM case, choosing the appropriate correlation points can be a difficult task \citep{Tilbrook2005}. A more often-used procedure, the domain integral method, which is an energy approach based on the $J$-integral \citep{Rice1968} that has been proved to be efficient for homogeneous materials, is used in this work.

From a numerical and computational perspective, one of the challenges concerns the need for examining the limiting case of a vanishing contour for the proper evaluation of the $J$-integral for crack tips in FGMs. This need stems from the fact that for some non-homogeneous materials and crack tip orientations, the integrand in the $J$-integral is not divergence free \citep{Chen2000}. As a result, an evaluation of the integral on open contours will exhibit path dependence. Thus, the standard $J$-integral along an integral path $\Gamma$ is defined as:

\begin{equation}
J = \int_\Gamma (W \delta_{1j} - \sigma_{ij}u_{i,1})n_j \,d\Gamma
\end{equation}

where $W$ is the strain energy density and $n_j$ is the outward normal to the path $\Gamma$. For a closed boundary $\Gamma = \Gamma_1 + \Gamma_A + \Gamma_B - \Gamma_0$ as shown in Fig. 5, the $J$-integral is formulated such that:

\begin{equation}
I = \oint_{\Gamma} (W \delta_{1j} - \sigma_{ij}u_{i,1})n_j \,d\Gamma = J_{\Gamma_1} + J_{\Gamma_A} + J_{\Gamma_B} - J_{\Gamma_0}
\end{equation}

Applying Gauss's divergence theorem gives:

\begin{equation}
I = \int_A [W_{,1} - (\sigma_{ij}u_{i,1})_{,j}] \,dA 
\end{equation}

where

\begin{equation}
W_{,1}=(\frac{1}{2} \sigma_{ij} \varepsilon_{ij})_{,1} = \sigma_{ij} \varepsilon_{ij,1} + \frac{1}{2} \varepsilon_{ij} D_{ijkl,1} \varepsilon_{kl}
\end{equation}

\begin{figure}
\centering
\includegraphics{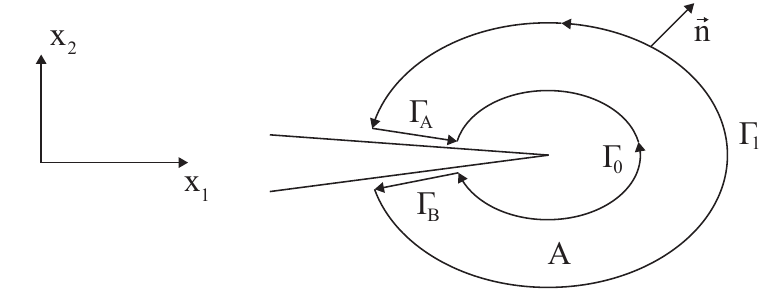}
\caption{J-integral Domain}
\label{fig:Fig5}
\end{figure}

Equations of equilibrium in the absence of body forces take the form $\sigma_{ij,i}=0$, and therefore:

\begin{equation}
\sigma_{ij} \varepsilon_{ij,1} = (\sigma_{ij} u_{i,1})_{,j}
\end{equation}

Substituting Eqs. (4) and (5) in Eq. (3) renders:

\begin{equation}
I = \int_A \frac{1}{2} \varepsilon_{ij} D_{ijkl,1} \varepsilon_{kl} \,dA 
\end{equation}

Along crack sides, $dx_2 = 0$ and the traction $t_i = \sigma_{ij} n_j$ is also zero. Consequently $J_{\Gamma_A}=J_{\Gamma_B}=0$ and thus:

\begin{equation}
I = J_{\Gamma_1} - J_{\Gamma_0}
\end{equation}

In an homogeneous material, since $D_{ijkl,1}=0$, $I=0$ and $J_{\Gamma_0}=J_{\Gamma_1}$, the $J$-integral is path independent. For FGMs, generally $D_{ijkl,1} \neq 0$, therefore $I \neq 0$ and $J_{\Gamma_0}\neq J_{\Gamma_1}$, whereby the $J$-integral is related to the integral path. When the material properties only vary along the $x_2$ axis, $D_{ijkl,1}=0$ and in consequence, for this case, the $J$-integral is still path independent.

Considering a smooth function $q$ which has he value of unity on $\Gamma_1$ and zero on $\Gamma_0$, the $J$-integral given in (1) can be written in terms of a closed boundary integral:

\begin{equation}
J = \oint_{\Gamma} (\sigma_{ij}u_{i,1} - W \delta_{1j})q n_j \,d\Gamma
\end{equation}

Applying the divergence theorem and assuming a Young's modulus variation along the $x_1$ axis, the corresponding expression for the domain integral is obtained:

\begin{equation}
J = \int_A (\sigma_{ij}u_{i,1} - W \delta_{1j})\,q_{,j}\,dA - \int_A W_{,1} q \, dA
\end{equation}

Being the derivative of $W$ under the second integral with respect to the coordinate $x_1$ in $E(x_1)$. So, comparing with the homogeneous case, the second integral is an additional term which represents the effect of non-homogeneity. But if the domain integral is evaluated in a region sufficiently small around the crack tip, the value of the second term in Eq. (9) involving the derivative of $W$ is very small, essentially negligible \citep{Gu1997}. The same conclusions can be extended to the Interaction Integral \citep{Shih1988} as it is calculated using the same procedure. Under mixed-mode conditions many finite element packages, including ABAQUS, provide an interaction integral method to compute the SIFs since it is not straightforward to do it from a known $J$-integral. The interaction integral is also used for the computation of the non-singular term of the $T$-stress.

Therefore, a sufficiently refined mesh near the crack tip was used in all the specimens analyzed in order to allow for comparison between all cases. Although the standard domain integral combined with a fine mesh allows to reach accurate results, this greatly increases the numerical costs. In order to exploit the advantage of the contour integral, which does not require a very fine mesh, one should use the above mentioned research codes, FGM-FRANC2D or WARP3D, which take into account the additional term of the Domain Integral necessary to maintain the path independence. Therefore, the computation of the derivative of the strain energy density with respect to the crack front (normal) direction is a pending task for future developments of commercial FE software.

In this work, the entire specimens were modeled using eight-noded quadrilateral plane-stress elements with reduced integration (CPS8R). An example of the mesh used in the simulations is shown in Fig. 6. The near-tip discretization consisted of a focused mesh centered at the crack tip and the inverse square root singularity of the strain field at the crack tip is obtained by collapsing quadrilateral elements into triangular elements and placing the mid-side nodes at quarter-points from the crack tip. Approximately 30000 elements were used in total. 

\begin{figure}[htbp]
\centering
\includegraphics{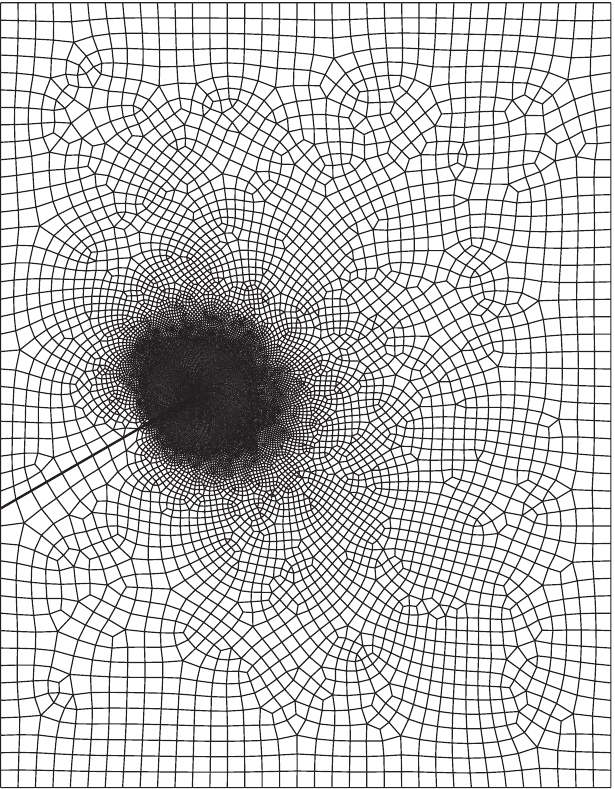}
\caption{Representative finite element mesh used for calculating fracture parameters}
\label{fig:Fig6}
\end{figure}

\subsubsection{Simulation of crack propagation}

Crack propagation was simulated by means of the X-FEM. The X-FEM is a numerical technique for modeling discontinuities by local enrichment functions in the area of interest. This method allows to follow crack paths independently of the finite element mesh; being this feature especially important for FGMs, since the gradation of the mechanical properties may lead to complex propagation paths also in simple symmetric tests \citep{Rousseau2000}. Cracks are modeled by means of the X-FEM using the cohesive segments method \citep{Remmers2008} implemented in ABAQUS software. This method is based on the principle of phantom nodes \citep{Song2006} and a traction-separation cohesive behavior.

In quasi-brittle materials a crack is usually assumed to grow when the tensile strength is attained at the current crack tip; propagation then occurs perpendicularly to the direction of the maximum in-plane principal stress. Therefore, crack initiation and propagation direction was predicted using the local criterion of the maximum principal stress (MPS). Based on the concept of local homogenization near the crack tip \citep{Gu1997b}, fracture criteria originally developed for homogeneous materials can be extended to non-homogeneous materials such as FGMs. The values of the initial failure stress and the subsequent fracture energy characterizing damage evolution are assigned taking into consideration the local failure properties recorded by \cite{Abanto-Bueno2006} (see Fig.1). Fracture energy variation was provided following the same procedure as with the elastic gradient implementation, although, based on the local propagation criteria established, the changes of fracture resistance as a function of the position only influence the amount of applied load necessary to propagate the crack. Notice that there is no effect of the load magnitude on the crack trajectory within the framework of linear elastic analysis. All the specimens were discretized with the same mesh density by means of approximately 450 four-noded quadrilateral plane-stress elements with reduced integration (CPS4R). An example of the mesh used in the simulations is shown in Fig. 7.

\begin{figure}[H]
\centering
\includegraphics{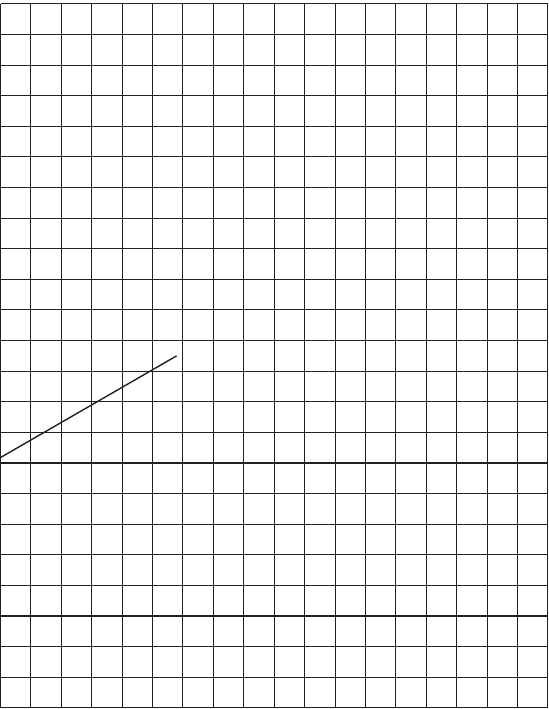}
\caption{Representative mesh used for predicting propagation paths of cracks by means of the X-FEM}
\label{fig:Fig7}
\end{figure}

\section{Results and discussion}
\label{}

\subsection{Crack initiation}

\subsubsection{Homogeneous Materials}

The base homogeneous material studied is an ECO sheet irradiated uniformly for 50 h under UV light. Its elastic properties were measured by \cite{Abanto-Bueno2006} as Young's modulus E=280 MPa and Poisson's ratio $\nu=0.45$. Geometry and dimensions are shown in Fig. 8.

\begin{figure}[htbp]
\centering
\includegraphics{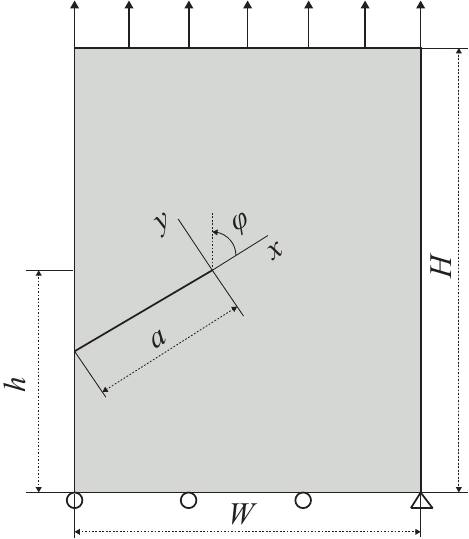}
\caption{Edge cracked specimen geometry for homogeneous material, $H$=90 mm, $W$=70 mm, $h$=45 mm, $a$=33 mm and $\varphi=\pi/3$}
\label{fig:Fig8}
\end{figure}

By means of the procedure outlined in the previous section, values of $K_I$, $K_{II}$ and $T$ corresponding to the experimentally recorded instant of crack initiation were calculated. Table 3 shows a comparison of the results obtained with the corresponding experimental values extracted by \cite{Abanto-Bueno2006} and the results obtained numerically by the same authors in collaboration with other researchers in a related work \citep{Oral2008}. Good agreement is observed, especially for the non-singular term of the $T$-stress, which is usually more difficult to obtain.

\begin{table*}[!ht]
\caption{Experimental and numerical results for $K_I$, $K_{II}$ and the $T$-stress for the homogeneous edge cracked specimen}
\centering
\begin{tabular}{c c c c c} 
\hline \hline
 \multicolumn{2}{c}{Results} & $K_I (M\!Pa\,m^{0.5})$ & $K_{II} (M\!Pa\,m^{0.5})$ & $T (M\!Pa)$\\
 \hline
 Expt. & \cite{Abanto-Bueno2006} & 0.903 & 0.245 & -0.784 \\
 \multirow{2}{*}{Num.} & \cite{Oral2008} & 0.793 & 0.212 & -0.992\\
 & Present & 0.812 & 0.291 & -0.647\\ 
 \hline \hline 
\end{tabular}
\label{tab:Table3}
\end{table*}

\begin{table*}[!ht]
\caption{Experimental and numerical results for $K_I$, $K_{II}$ and the $T$-stress for the FGM edge cracked specimens}
\centering
\begin{tabular}{c c c c c c} 
\hline \hline
 \multicolumn{2}{c}{Results} & Case & $K_I (M\!Pa\,m^{0.5})$ & $K_{II} (M\!Pa\,m^{0.5})$ & $T (M\!Pa)$\\
 \hline
 Expt. & \cite{Abanto-Bueno2006}& & 0.554 & 0.039 & -4.272\\
 \multirow{2}{*}{Num.} & \cite{Oral2008}& I & 0.551 & -0.022 & -2.149\\
 & Present & & 0.599 & -0.018 & -1.526\\ 
 \\
 Expt. & \cite{Abanto-Bueno2006} & & 0.755 & 0.179 & -0.069\\
 \multirow{2}{*}{Num.} & \cite{Oral2008}& II & 0.722 & 0.204 & -0.673\\
 & Present & & 0.734 & 0.257 & -0.539\\
 \\
 Expt. & \cite{Abanto-Bueno2006} & & 0.969 & 0.224 & -0.930\\
 \multirow{2}{*}{Num.} & \cite{Oral2008}& III & 0.878 & 0.230 & -0.870\\
 & Present & & 0.908 & 0.304 & -0.763\\
 \hline \hline 
\end{tabular}
\label{tab:Table4}
\end{table*}
\vspace{10pt}

\subsubsection{Functionally Graded Materials}

Young's modulus distribution for the three graded specimens is shown in Fig. 2. In all cases, a constant value of 0.45 is considered for the Poisson's ratio, in accordance with the assumption of the experimental work \cite{Abanto-Bueno2006}. As described in the previous section, material property variation is implemented via a user subroutine USDFLD although in the present isothermal analysis, where the Poisson's ratio is constant and a very fine mesh is used, results agree with those obtained using Rousseau and Tippur's technique and therefore verify the subroutine implementation. Computed values of fracture parameters are compared with available experimental and numerical results in Table 4.

As in the homogeneous case, results agree reasonably well with the available experimental data, although there are some differences that need to be analyzed. The greatest discrepancies arise in the $T$-stress that, being a second-order term, is more difficult to extract, both numerically and experimentally. Besides, analyzed geometries bring great changes in its value, which is also affected by the direction of the material property variation. All the results, and especially the $T$-stress, are affected by the method chosen to extract them. As mentioned before, Abanto-Bueno and Lambros, in both their experimental \citep{Abanto-Bueno2006} and their related numerical work \citep{Oral2008} fit the experimentally measured or numerically computed displacements in the asymptotic displacement equation to extract the value of the fracture parameters, while in this work these are computed by means of the Domain Integral without the need of any postprocessing. Also, due to the limitations of the DIC method, experimentally measured displacements exclude a rectangular region around the crack tip. Whereas in the numerical model data as close to the crack tip as possible are considered, in accordance with the local homogenization criterion.

In the first graded specimen, a negative value of $K_{II}$ is obtained, in contraposition with the positive value extracted from the experiments. The value is relatively small and whence the difference is minimal, but the sign affects the direction of crack kinking. The crack growth analysis conducted by \cite{Abanto-Bueno2006} shows a positive kink direction and therefore, the negative sign of $K_{II}$ obtained in the numerical simulation correctly predicts the positive sign of subsequent crack kinking.

Given the unavoidable experimental error and taking into consideration that differences found were duly justified, results show that finite element calculations allow to obtain fracture parameters for the precise instant of crack initiation with reasonable accuracy.

\subsection{Sensitivity analysis}

\subsubsection{Effect of an accurate fit of the local material properties in the crack-tip}

Among other factors, fracture parameters in non-homogeneous materials depend on the local properties at the crack tip. In order to quantify the influence of an accurate fit of the measured elastic properties in the vicinity of the crack, the value of the experimental point that characterizes the local value of the Young's modulus in the crack tip is slightly modified (see Table 2 and Fig. 3).

\begin{table}[h]
\caption{Percentage differences between each numerical curve fit of the experimental material properties and the original one}
\centering
\begin{tabular}{c c c c} 
\hline \hline
 & $FGMI$ (\%) & $FGMII$ (\%) & $FGMIII$ (\%) \\
 \hline
 -10\% & 0.0145\% & 0.0219\% & 0.0131\% \\
 -5\% & 0.0037\% & 0.0055\% & 0.0033\% \\
 +5\% & 0.0037\% & 0.0055\% & 0.0033\% \\
 +10\% & 0.0149\% & 0.0219\% & 0.0131\% \\
 \hline \hline 
\end{tabular}
\label{tab:Table5}
\end{table}

\begin{table*}[bp]
\caption{Numerical results for $K_I$, $K_{II}$ and $T$-stress for each test point considered. Values of FITs in $M\!Pa\,m^{0.5}$ and $T$-stress in $M\!Pa$ - FGMI}
\centering
\begin{tabular}{c c c c} 
\hline \hline
Test points & $K_I$ (\%) & $K_{II}$ (\%) & $T$ (\%) \\
 \hline
 -10\% & 0.591 (1.34\%) & -0.017 (5.55\%)& -1.509 (1.11\%)\\
 -5\% & 0.595 (0.67\%)& -0.017 (5.55\%)& -1.517 (0.59\%)\\
 Expt. data & 0.599 & -0.018 & -1.526\\
 +5\% & 0.603 (0.67\%)& -0.017 (5.55\%)& -1.535 (0.59\%)\\
 +10\% & 0.608 (1.50\%)& -0.017 (5.55\%)& -1.543 (1.11\%)\\
 \hline
 Expt. Results & 0.554 & 0.039 & -4.272\\
 \hline \hline 
\end{tabular}
\label{tab:Table6}
\end{table*}

As expected and it can be seen in Fig. 3, a small change in the value of one of the nearly 20 experimental points fitted does not bring many differences between the different least square fits derived from the changes proposed. In order to quantify these deviations, the average difference along the specimen width between the numerical fits emerged from the proposed changes and the numerical fit of the original experimental data is calculated in a percentage scale by means of the $L_2$ norm. Results obtained from this relationship are shown in Table 5.

Fracture parameters computed for each curve derived from these changes are presented in Tables 6 to 8. Percentage differences between the calculated values and the values of the fracture parameters corresponding to the numerical fit of the original experimental data are also shown for comparison purposes.

\begin{table*}[bp]
\caption{Numerical results for $K_I$, $K_{II}$ and $T$-stress for each test point considered. Values of FITs in $M\!Pa\,m^{0.5}$ and $T$-stress in $M\!Pa$ - FGMII}
\centering
\begin{tabular}{c c c c} 
\hline \hline
Test points & $K_I$ (\%) & $K_{II}$ (\%) & $T$ (\%) \\
 \hline
 -10\% & 0.720 (1.91\%)& 0.252 (1.95\%)& -0.527 (2.23\%)\\
 -5\% & 0.727 (0.95\%)& 0.255 (0.78\%)& -0.533 (1.11\%)\\
 Expt. data & 0.734 & 0.257 & -0.539 \\
 +5\% & 0.741 (0.68\%)& 0.260 (1.17\%)& -0.544 (0.93\%)\\
 +10\% & 0.747 (1.77\%)& 0.263 (2.33\%)& -0.550 (2.04\%)\\
 \hline
 Expt. Results & 0.755 & 0.179 & -0.069\\
 \hline \hline 
\end{tabular}
\label{tab:Table7}
\end{table*}

\begin{table*}[htbp]
\caption{Numerical results for $K_I$, $K_{II}$ and $T$-stress for each test point considered. Values of FITs in $M\!Pa\,m^{0.5}$ and $T$-stress in $M\!Pa$ - FGMIII}
\centering
\begin{tabular}{c c c c} 
\hline \hline
Test points & $K_I$ (\%) & $K_{II}$ (\%) & $T$ (\%) \\
 \hline
 -10\% & 0.895 (1.43\%)& 0.298 (1.97\%)& -0.759 (0.52\%)\\
 -5\% & 0.902 (0.66\%)& 0.301 (0.99\%)& -0.761 (0.26\%)\\
 Expt. data & 0.908 & 0.304 & -0.763\\
 +5\% & 0.914 (0.66\%)& 0.306 (0.66\%)& -0.764 (0.13\%)\\
 +10\% & 0.920 (1.32\%)& 0.308 (1.32\%)& -0.766 (0.39\%)\\
 \hline
 Expt. Results & 0.969 & 0.224 & -0.930\\
 \hline \hline 
\end{tabular}
\label{tab:Table8}
\end{table*}

Even though the differences between the numerical fits that emerged from the proposed changes are extremely small in the three graded specimens, very significant discrepancies, between 30 and 1000 times the differences between the numerical curve fits, can be found among the values of the fracture parameters. As a result more attention should be paid to an accurate fit of the experimental elastic properties in the vicinity of the crack.

\subsubsection{Influence of the shape of the numerical curve fit}

Besides the local properties in the crack tip, the shape of the material gradient could also influence the results. In order to evaluate the effect of an accurate fit of the elastic gradient profile, several polynomial curve fits of different orders are considered, containing all of them the experimental data characterizing the local property values in the crack tip, as seen in Fig. 4. Computed values for each curve fit of the SIFs and the $T$-stress for each graded specimen are presented in Tables 9 to 11. Experimental results of \cite{Abanto-Bueno2006} are also shown for comparison.

\begin{table*}[htbp]
\caption{Numerical results for $K_I$, $K_{II}$ and $T$-stress for each numerical fit considered - FGMI}
\centering
\begin{tabular}{c c c c} 
\hline \hline
Curve fit & $K_I (M\!Pa\,m^{0.5})$ & $K_{II} (M\!Pa\,m^{0.5})$ & $T (M\!Pa)$ \\
 \hline
 Pol. order 3 & 0.584 & -0.016 & -1.499\\
 Pol. order 4 & 0.599 & -0.018 & -1.526\\
 Pol. order 9 & 0.601 & -0.018 & -1.523\\
  \hline
 Expt. Results & 0.554 & 0.039 & -4.272\\
 \hline \hline 
\end{tabular}
\label{tab:Table9}
\end{table*}
\vspace{10pt}

\begin{table*}[htbp]
\caption{Numerical results for $K_I$, $K_{II}$ and $T$-stress for each numerical fit considered - FGMII}
\centering
\begin{tabular}{c c c c} 
\hline \hline
Curve fit & $K_I (M\!Pa\,m^{0.5})$ & $K_{II} (M\!Pa\,m^{0.5})$ & $T (M\!Pa)$ \\
 \hline
 Pol. order 3 & 0.745 & 0.261 & -0.539\\
 Pol. order 5 & 0.742 & 0.259 & -0.525\\
 Pol. order 9 & 0.743 & 0.260 & -0.533\\
  \hline
 Expt. Results & 0.755 & 0.179 & -0.069\\
 \hline \hline 
\end{tabular}
\label{tab:Table10}
\end{table*}
\vspace{10pt}

\begin{table*}[htbp]
\caption{Numerical results for $K_I$, $K_{II}$ and $T$-stress for each numerical fit considered - FGMIII}
\centering
\begin{tabular}{c c c c} 
\hline \hline
Curve fit & $K_I (M\!Pa\,m^{0.5})$ & $K_{II} (M\!Pa\,m^{0.5})$ & $T (M\!Pa)$ \\
 \hline
 Pol. order 3 & 0.975 & 0.329 & -0.768\\
 Pol. order 5 & 0.972 & 0.327 & -0.774\\
 Pol. order 9 & 0.965 & 0.322 & -0.772\\
  \hline
 Expt. Results & 0.969 & 0.224 & -0.930\\
 \hline \hline 
\end{tabular}
\label{tab:Table11}
\end{table*}
\vspace{10pt}

Tables 9 to 11 show sensitivity in all the fracture parameters to the different curve fit considered in the three specimens evaluated. Although changes in the order of the polynomial function of the least squares fit do not affect the SIFs and the $T$-stress as much as the variation of the local elastic properties at the crack tip, results prove the influence of the variation of the elastic properties by itself in the calculation of fracture parameters. Therefore, significant differences between the curve fit and the experimentally measured elastic properties can bring inaccuracies in the calculations. Consequently, the numerical fit must be as accurate as possible in the vicinity of the crack, but without neglecting the relevance of a rigorous fit of the rest of experimental data. In order to solve this problem a new method is proposed: a point to point lineal fit.

\subsubsection{Point to point lineal fit}

\begin{figure*}[!ht]
    \subfloat[\label{subfig-1:Fig9a}]{%
      \includegraphics[width=0.45\textwidth]{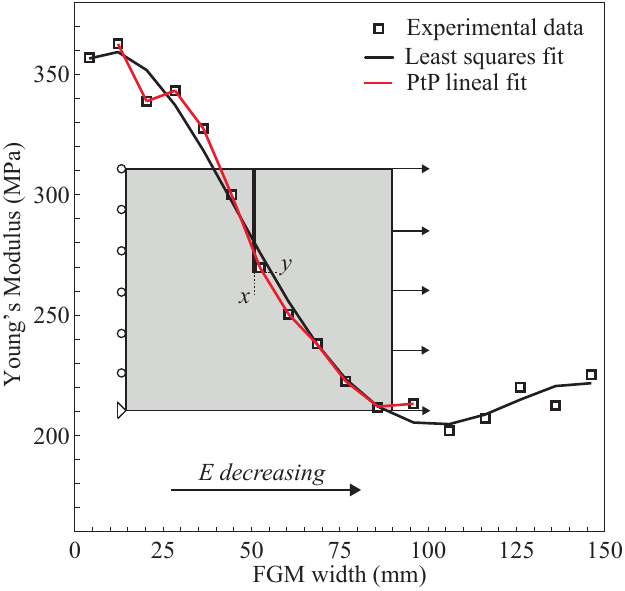}
    }
    \hfill
    \subfloat[\label{subfig-2:Fig9b}]{%
      \includegraphics[width=0.45\textwidth]{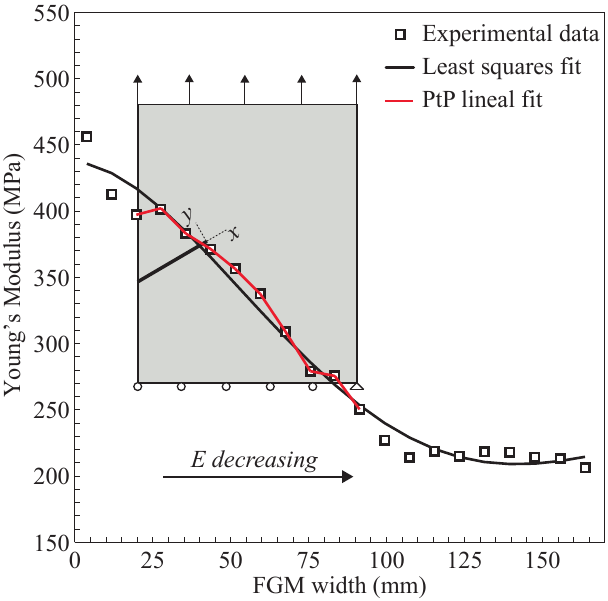}
    }
    
    \centering
    \subfloat[\label{subfig-2:Fig9c}]{%
      \includegraphics[width=0.45\textwidth]{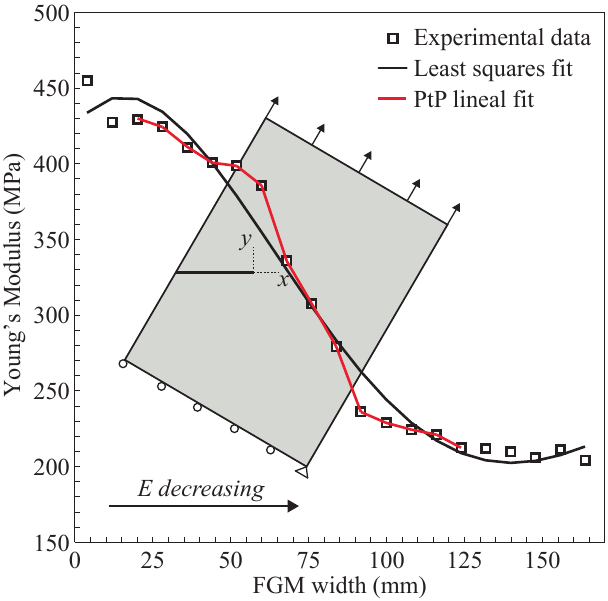}
    }
    \caption{Numerical fit of the material property variation of \textbf{(a)} FGMI, \textbf{(b)} FGMII and \textbf{(c)} FGMIII}
    \label{fig:Fig9}
\end{figure*}

With the aim of removing the differences between the data fit and the experimental data arises the possibility to implement directly in the numerical model the values of the Young's Modulus extracted from the experiments. In order to do so, elastic properties are designated as a function of a field variable (or temperature) and each one of the experimental registered values of the Young's Modulus in the specimen is defined with an assigned value of the field variable that depends on its \textit{position in the sequence of experimental data}. Afterwards, the specimen is provided with a lineal distribution of the field variable that allocates the points depending on its position in the specimen. The finite element software assigns by default the value of the elastic properties in places where they are not defined considering a linear variation between the nearest two points registered. This results in a linear change between the data points as is indicated by the red line in Fig. 9 for the three specimens evaluated. The polynomial curve fit used until now is also presented with the aim of establishing a comparison.

In order to be accurate, this method is limited by the fact that the experimental data needs to be separated by a distance as similar as possible. However, this is usually not an obstacle since when property variation is determined directly from experiment, as in most cases, measurements are performed on points of the material containing a constant separation from each other with the aim of correctly record the variation of the material properties. Through the same procedure that has been used so far, fracture parameters are calculated and shown in Table 12. Experimental data from \cite{Abanto-Bueno2006} and the results from the common used polynomial fit are also shown.

\begin{table*}[htbp]
\caption{Numerical results for $K_I$,$K_{II}$ and $T$-stress for each numerical fit considered}
\centering
\begin{tabular}{c c c c c} 
\hline \hline
Procedure & Case & $K_I (M\!Pa\,m^{0.5})$ & $K_{II} (M\!Pa\,m^{0.5})$ & $T (M\!Pa)$ \\
 \hline
 Experimental & & 0.554 & 0.039 & -4.272 \\
 Polynomial fit & I & 0.599 & -0.018 & -1.526\\
 PtP lineal fit & & 0.602 & -0.018 & -1.529\\
 \\
 Experimental & & 0.755 & 0.179 & -0.069 \\
 Polynomial fit & II & 0.734 & 0.257 & -0.539 \\
 PtP lineal fit & & 0.718 & 0.251 & -0.517\\
 \\
 Experimental & & 0.969 & 0.224 & -0.930 \\
 Polynomial fit & III & 0.908 & 0.304 & -0.763\\
 PtP lineal fit & & 0.935 & 0.296 & -0.942\\
 \hline \hline 
\end{tabular}
\label{tab:Table12}
\end{table*}

As seen in Table 12, even though the experimental error is unknown, results improve when using the point to point lineal fit. Especially in the third specimen, where the polynomial function loses accuracy when fitting the experimental data in the vicinity of the crack (Fig. 9).

\subsection{Crack growth}

Crack propagation trajectories of the three specimens were calculated based on the X-FEM and the local fracture criterion of the MPS. Results are compared with available experimental data from \cite{Abanto-Bueno2006} in Fig. 10 and the performance of local fracture criterion in the prediction of the crack propagation path is analyzed.

As seen in Fig. 10 numerical prediction approaches reasonably well the experimental data. In the case of the first graded specimen, in both the numerical (Fig. 10 (b)) and the experimental results (Fig. 10 (a)) it can be appreciated that the crack propagation path deviates from the natural trajectory of the first mode of fracture. This is due to the fact that the crack is initially orientated perpendicular to the material gradient, creating an asymmetric stress field around the crack tip that induces mixed-mode loading causing crack deflection. As expected, the crack propagated towards the more compliant part. This is understood to occur as it results in greater release of elastic potential energy.

\begin{figure*}[!ht]
    \subfloat[\label{subfig-1:Fig10a}]{%
      \includegraphics[width=0.4\textwidth]{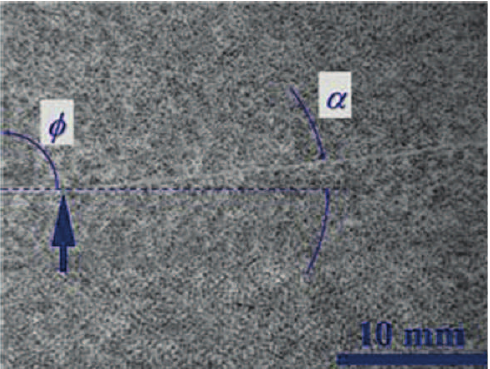}
    }
    \hfill
    \subfloat[\label{subfig-2:Fig10b}]{%
      \includegraphics[width=0.4\textwidth]{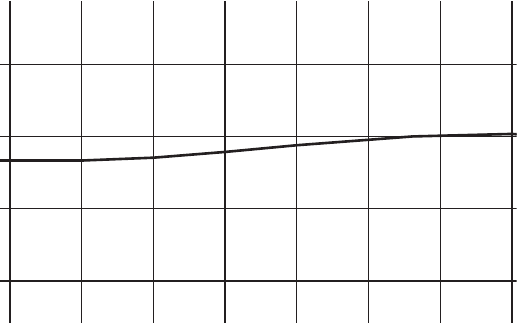}
    }
    
    \subfloat[\label{subfig-2:Fig10c}]{%
      \includegraphics[width=0.4\textwidth]{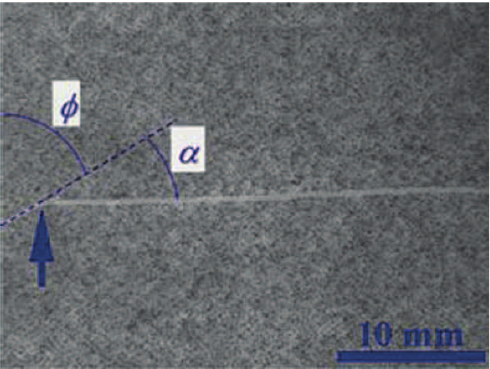}
    }
    \hfill
    \subfloat[\label{subfig-2:Fig10d}]{%
      \includegraphics[width=0.4\textwidth]{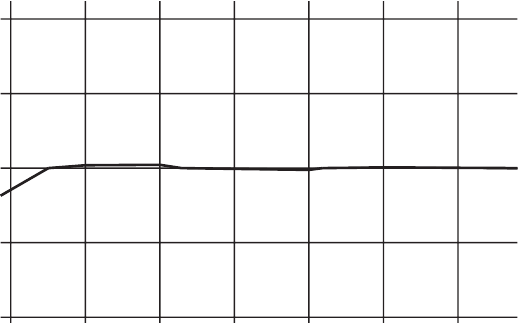}
    }
    
    \subfloat[\label{subfig-1:Fig10e}]{%
      \includegraphics[width=0.4\textwidth]{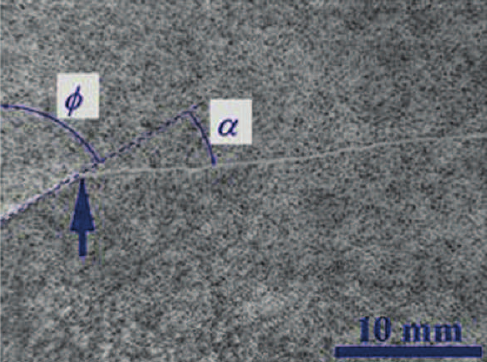}
    }
    \hfill
    \subfloat[\label{subfig-2:Fig10f}]{%
      \includegraphics[width=0.4\textwidth]{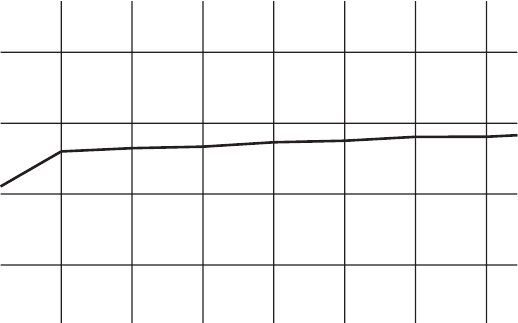}
    }
    \caption{Comparison of crack trajectories obtained experimentally by \cite{Abanto-Bueno2006} and the present X-FEM simulation.}
    \label{fig:Fig10}
\end{figure*}

Crack propagation path for FGMII is shown in Fig. 10 (c) and (d). Under the same loading conditions in an homogeneous specimen, the crack propagation path would register a small initial deviation, due to the asymmetry of the loading within the initial crack orientation, to grow then according to the trajectory distinctive of the first mode of fracture. So that the initial asymmetry in the exact moment of the onset of crack propagation due to the position of the crack within the direction of mechanical property variation and the external loading translates over the trajectory exclusively in a small initial deflection, that comes across very well in the numerical simulation and which is also referred in Abanto-Bueno and Lambros experimental work \citep{Abanto-Bueno2006}. After this initial deviation, the crack grows following the propagation path distinctive to the first mode of fracture since is oriented parallel to the material property variation
 
In FGMIII at the precise instant of crack initiation there is a mixed-mode fracture induced by the asymmetry of the loading within the crack position. However, in an homogeneous specimen with the same loading conditions, after an initial deviation the crack should grow according to the idiosyncratic trajectory of the first mode of fracture. As seen in Fig. 10 (e) and (f), this is not the case, as once the crack trajectory tries to get horizontal, forms an angle with the gradient direction and therefore, the asymmetric stress field around the crack tip induces mixed-mode loading causing crack deflection.
 
Although numerical predictions show reasonably good agreement with the experimental data, a close observation of the results of the first and third cases shows that the deviation of the trajectory from the crack propagation path corresponding to the first mode of fracture is slightly higher in the experimental case. This may be due to many factors and it is not surprising that there are some discrepancies between predicted and observed deflection trajectories, given the stochastic nature of crack initiation from a notch and the possible influences of the microstructure. In fact, assumed local homogenization criteria is expected to work well only for ideally brittle non-homogeneous materials where the toughness is taken to be independent of direction at a fixed point \citep{Gu1997b}. However, in a FGM specimen, due to its change in material composition, this is often not the case, and therefore the material toughness variation as a function of the position cannot be neglected. So that, in order to develop a more accurate crack propagation criterion for non-homogeneous materials, this should be stated as:

\begin{equation}
\frac{\partial}{\partial \alpha}(G-G^R)=0
\end{equation}

Where $G$ is the energy release rate and $G^R$ is the critical value of the energy release rate according to which the crack starts to propagate, the toughness of the FGM. Although in the specimens evaluated in this work the spatial variation of the fracture resistance is unknown, based on the variation of the failure properties registered by \cite{Abanto-Bueno2006} and shown in Fig. 1, one could expect that it would increase the deviation predicted based on the effect of the elastic gradient of the material, as it varies in the same direction in the all the specimens evaluated.

Nevertheless, very little work has been published on crack propagation in graded regions. Crack trajectory becomes more complex after deviating from a straight path and therefore, appropriate models for the continued propagation of cracks forming an angle with the material gradient will require a fracture criterion which incorporates the effects of variations in local mechanical properties, crack-tip toughness, bridging and residual stresses, as well as changing crack shape \citep{Tilbrook2005b}. To the authors' knowledge, this fracture criterion has not been developed yet.

\section{Concluding remarks}
\label{}

In this work, performance of numerical tools to study the structural integrity of advanced materials is evaluated. Crack initiation and growth in planar FGMs is investigated by means of the well-known finite element package ABAQUS.

With the aim of overcoming the limitations of Rousseau and Tippur's technique \citep{Rousseau2000}, material gradient was implemented by means of a user subroutine USDFLD and its template is provided in the Appendix in order to facilitate the work of other other researchers and practitioners. When computing fracture parameters in a precise instant of the fracture process, commercially available finite element software has proven to obtain accurate results, although the use of the before mentioned research codes, FGM-FRANC2D or WARP3D, is recommended when the crack is not perpendicular to the direction of the material property variation in order to reduce CPU time.

However, when predicting crack propagation paths, available finite element codes can only get accurate results by means of local crack propagation criteria if the elastic gradient is the dominant influence on crack propagation. To develop a general purpose formulation and implementation for predicting crack trajectories in FGMs a new fracture criterion that incorporates the influence of variations in crack-tip toughness as well as other mechanical effects, is needed.

In addition, this paper emphasizes that more attention needs to be paid to the numerical fit of the experimental data characterizing the change in the material properties, and a new method is proposed to improve the accuracy of the results. The commonly used criterion based on a general approximation of the overall experimental data could introduce inaccuracies in the calculations within a fracture analysis, as small changes in the local elastic properties at the crack-tip have been shown to have significant influence in the values of the fracture parameters. Furthermore, an accurate fit of the shape of the material gradient is also relevant, as even by itself it influences the fracture parameters. A point to point lineal fit has proven to reach more accurate results.

\begin{acknowledgements}
The authors gratefully acknowledge the financial support from the Ministry of Science and Innovation of Spain through the grant DPI2010.21590.CO2.01.
\end{acknowledgements}

\section*{Appendix A. User subroutine USDFLD FGMII}
\label{}
\noindent
\texttt{C User subroutine for the implementation of\\
C a continuous variation of the material\\
C elastic properties between integration \\
C points.\\\\
C Code layout is provided the simplest way \\
C possible and comments are included\\
C conveniently with the aim that it can serve\\
C as a template. Fortran statements are \\
C placed at the beginning of each row based\\
C on text formatting, make sure they are\\
C placed in the appropriate column before\\
C compiling the code.\\\\
C The user subroutine USDFLD will be called \\
C at all the integration points of the model \\ 
C for which the material definition includes\\
C user-defined field variables. As a matter\\
C of example, the elastic gradient profile of\\
C the second specimen evaluated is \\
C implemented (FGMII).\\\\
SUBROUTINE USDFLD(FIELD,STATEV,PNEWDT,DIRECT,\\
1 T,CELENT,TIME,DTIME,CMNAME,ORNAME,NFIELD,\\
2 NSTATV,NOEL,NPT,LAYER,KSPT,KSTEP,KINC,NDI,\\
3 NSHR,COORD,JMAC,JMATYP,MATLAYO,LACCFLA)\\
C\\
INCLUDE 'ABA\_PARAM.INC'\\
C\\
CHARACTER*80 CMNAME,ORNAME\\
CHARACTER*3  FLGRAY(15)\\
DIMENSION FIELD(NFIELD),STATEV(NSTATV),\\
1 DIRECT(3,3),T(3,3),TIME(2)\\
DIMENSION ARRAY(15),JARRAY(15),JMAC(*),\\
1 JMATYP(*),COORD(*)\\\\
C In accordance with the procedure adopted \\
C by ABAQUS, variables are defined implicitly\\
C and therefore, its type is determined by \\
C its first letter.\\\\
C X and Y are variables of the type real \\
C that store the coordinates of the \\
C corresponding integration point relative to\\
C the horizontal and vertical axis, \\
C respectively.\\\\
X=COORD(1)\\
Y=COORD(2)\\\\
C As seen in Fig. 2, the material elastic\\ 
C properties are assumed to vary according\\
C to a fourth order polynomial function. Note\\
C that the origin of the coordinate axes has\\
C been placed following the notation adopted\\ 
C in Fig. 2.\\
C It is also important to note that, in the\\ 
C material definition, a direct relation \\ 
C between the Young's Modulus (MPa) and the\\
C FIELD1 variable is designated so that both\\ 
C variables adopt the same values and \\
C therefore vary equally.\\\\
FIELD(1)= - 916659.61963*X**4 +\\
1 410881.971401*X**3 - 49875.3879242*X**2 -\\
2 185.636233064*X + 437.202232184\\\\
RETURN\\
END\\\\
C In this case, a UFIELD subroutine is used \\ 
C in conjunction with the USDFLD subroutine \\
C defined above in order to implement the \\
C material elastic properties variation not \\
C only in the integration points of the \\
C model, but also in the nodes. This is only\\ 
C necessary when computing the stress \\
C intensity factors in a fracture mechanics \\
C analysis, since the elastic properties \\
C in the crack tip must be defined. The\\
C terminology used is the same as in the \\ 
C USDFLD subroutine.\\\\
SUBROUTINE UFIELD(FIELD,KFIELD,NSECPT,KSTEP,\\
1 KINC,TIME,NODE,COORDS,TEMP,DTEMP,NFIELD)\\
C\\
INCLUDE 'ABA\_PARAM.INC'\\
C\\
DIMENSION FIELD(NSECPT,NFIELD), TIME(2), \\
1 COORDS(3), TEMP(NSECPT), DTEMP(NSECPT)\\
C\\
X=COORDS(1)\\
Y=COORDS(2)\\\\
FIELD(1,1)= -916659.61963*X**4 + \\
1 410881.971401*X**3 - 49875.3879242*X**2 -\\
2 185.636233064*X + 437.202232184\\\\
RETURN\\
END}

% BibTeX users please use one of
\bibliographystyle{spbasic}      % basic style, author-year citations
\bibliography{bibliography}   % name your BibTeX data base

\end{document}